# Zero-Point Renormalization Reshapes the Shift Current: The Role of Electron-Phonon Coupling

Yang Li[1,2,*], Zuzhang Lin[3,4,5], Zhikuan Wang[2], Chengzhi Wu[2], Zeyu Jiang[2], and Bing Huang[2,6,*]

[1]*School of Mathematics and Physics, University of Science and Technology Beijing, Beijing 100083, China*

[2]*Beijing Computational Science Research Center, Beijing 100193, China*

[3]*New Cornerstone Science Lab, Department of Physics, The University of Hong Kong, Hong Kong, China*

[4]*HK Institute of Quantum Science & Technology, The University of Hong Kong, Hong Kong, China*

[5]*State Key Laboratory of Optical Quantum Materials, The University of Hong Kong, Hong Kong, China*

[6]*School of Physics and Astronomy, Beijing Normal University, Beijing 100875, China*

**Correspondence: liyang1991@ustb.edu.cn, bing.huang@csrc.ac.cn*

## Abstract

Shift current (SC) drives the bulk photovoltaic effect, holding promise for near-instantaneous, low-dissipation photodetection and energy conversion. The current understanding of SC relies primarily on a single-particle picture, in which the SC is determined purely by band dispersion. While many-body effects from directly excited quasiparticles, such as excitons, magnons and phonons, have been explored, the renormalization of electronic states by quantum virtual-phonon coupling — inherent in essentially all solids and particularly decisive for transport — remains unaddressed. Here, using first-principles calculations, we identify this missing channel — band-renormalized shift current (BRSC) — in monolayer $MoS_2$. We find that the zero-point renormalization alone, without any polarization-carrying photoexcited phonons and with barely any alteration of the band dispersion, dramatically reshapes the SC spectrum: emergent low-energy peaks arise from phonon-assisted optical transitions between momentum-distinct states, with a pronounced redshift of higher-energy features. A mere 1.0% strain sharply tunes the BRSC, whereas raising the temperature enhances it. Our findings overturn the band-structure-only paradigm, establishing electron-phonon coupling as an indispensable ingredient for nonlinear photocurrent and opening a strain-thermal route for engineering next-generation optoelectronic responses.

*Introduction.* Shift current (SC) — a nonlinear photocurrent arising in non-centrosymmetric materials under illumination — is the primary origin of the bulk photovoltaic effect (BPVE) [1–5]. Physically, SC originates from the shift of carrier wave packets upon interband optical transitions, quantified by the shift vector constructed from Berry connections [6–10]. The quantum-geometric origin of SC and its promise for near-instantaneous, junction-free photodetection and energy conversion [3–5] have driven rapid expansion across a broad range of systems, including ferroelectrics [1, 2], topological materials [11, 12], van der Waals compounds [13–18], and magnetic systems [19–24]. Theoretically, SC is understood within the single-particle framework established more than four decades ago [6–8]: the centers of carrier wave packets shift during vertical optical excitation, producing a direct current in real space [Figs. 1(a) and 1(c)], and the SC conductivity — an integral of the shift vector weighted by linear optical absorption — is dictated entirely by the band structure.

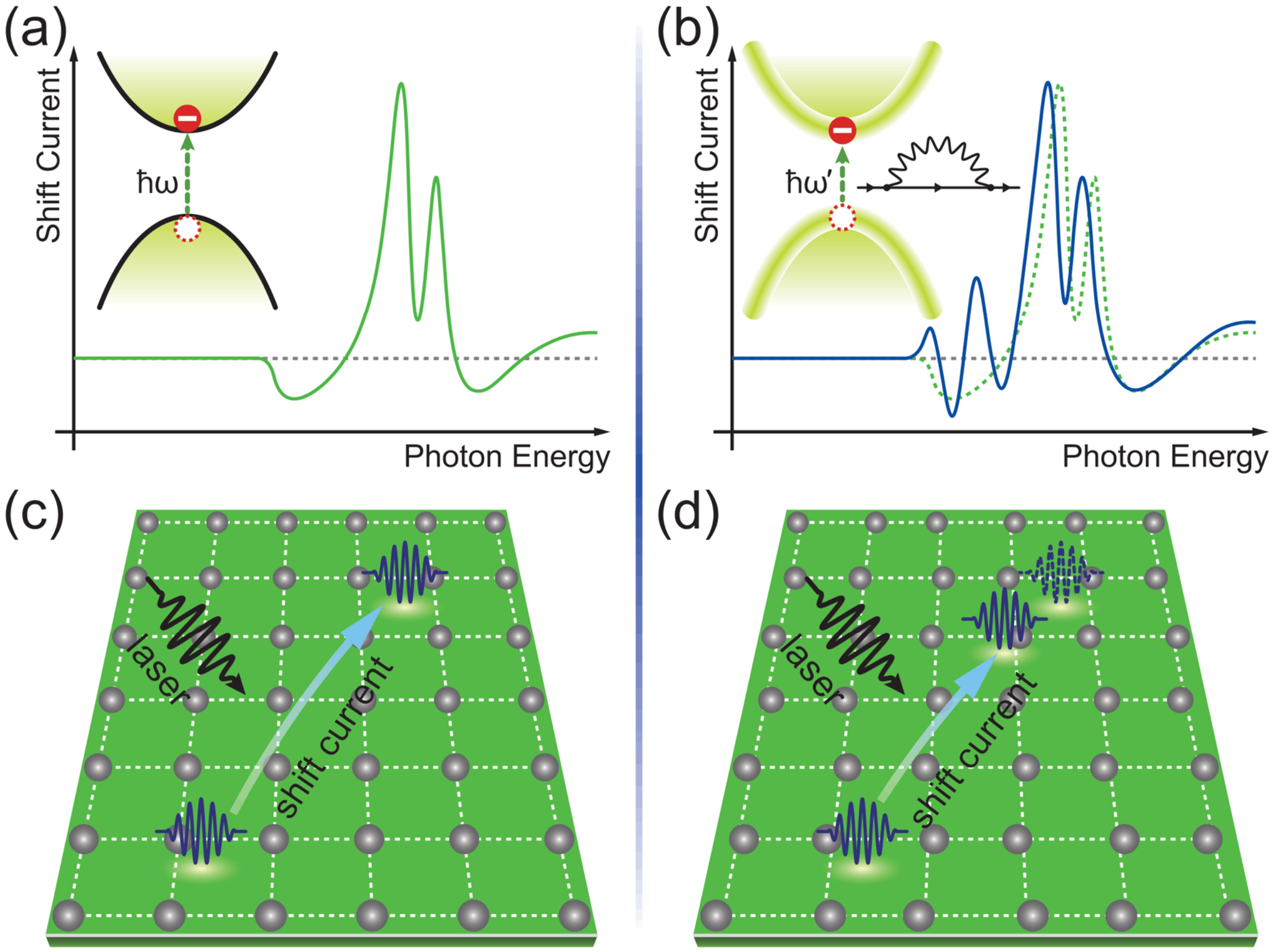

**FIG. 1.** Schematic illustration of the shift current (SC) and the band-renormalized shift current (BRSC). Left column [(a), (c)]: Single-particle picture. (a) Only vertical optical transitions occur, producing the conventional SC spectrum. (c) Photoexcitation displaces the carrier wave packets, generating a direct real-space current. Right column [(b), (d)]: Extended picture with electron-phonon coupling (EPC). (b) Band renormalization and broadening enable phonon-assisted transitions between momentum-distinct

states, giving rise to emergent peaks in the BRSC spectrum; the green dashed curve shows the spectrum without EPC for comparison. (d) The wave-packet shift is modified accordingly (solid blue curve), deviating from its single-particle counterpart (dashed blue curve). Black squiggly lines denote incident photons.

However, this framework omits a ubiquitous many-body interaction: electron-phonon coupling (EPC). In contrast to purely optical processes such as second-harmonic generation, photocurrent is a transport phenomenon and is therefore inherently sensitive to EPC. Yet existing treatments of EPC in the SC context are limited to classical thermal lattice vibrations [25], while recent many-body extensions of the single-particle picture have focused on directly excited polarization-carrying real quasiparticles such as excitons [26–30], magnon [31, 32] and phonons [33–35]. The EPC-driven renormalization of the electronic states themselves, induced by quantum virtual-phonon coupling — present even at zero temperature — remains unexplored. Generally, EPC renormalizes electronic bands and broadens optical spectra [36]. This renormalization modifies the real-space displacement of photoexcited wave packets, giving rise to what we term the *band-renormalized shift current* (BRSC) [Figs. 1(b) and 1(d)]. Moreover, many BPVE applications operate at finite temperatures and experience light-induced heating, both of which can markedly affect the SC through EPC-governed quasiparticle evolution. A comprehensive understanding of the BRSC is thus essential for predicting realistic device performance.

Monolayer $MoS_2$, with its simple band structure and strong nonlinear optical (NLO) response, provides an ideal platform for isolating this many-body effect. In this Article, we employ density functional theory (DFT) calculations and finite-difference method (see Sec. I in the Supplemental Material (SI) [37]) to study the EPC-induced BRSC in monolayer $MoS_2$. We find that zero-point renormalization (ZPR) at zero temperature — although barely altering the band dispersion — suffices to reshape the SC spectrum: emergent low-energy peaks arise because electron-phonon scattering enables otherwise forbidden optical transitions between states with different momenta, while higher-energy peaks exhibit a pronounced redshift driven by multiple competing phonon-assisted channels. Modest biaxial strain (~1.0%) sharply modulates the BRSC by shifting the transition energies of the momentum-distinct states involved, and raising the temperature enhances the BRSC through the increased phonon population. The mechanism established here could be material-independent and applies generally to nonlinear photocurrents in non-centrosymmetric systems.

*Results.* The common structure of monolayer $MoS_2$, *i.e.*, the 2H structure, belongs to the $P\bar{6}m2$ space group, with a three-fold rotational axis, three mirror planes inclined at 60° to each other, and one mirror plane perpendicular to the out-of-plane direction (see Sec. II in the SI) [37]. The absence of inversion symmetry permits second-order NLO responses [38], and the $D_{3h}$ symmetry leaves a single non-zero and independent susceptibility component, $\chi^{(2)}_{yxx} = -\chi^{(2)}_{yyy} = \chi^{(2)}_{xyx} = \chi^{(2)}_{xxy}$ [39]. We therefore focus on the SC conductivity $\sigma^{(2)}_{yxx}$ in the following.

Figure 2(a) displays the calculated band structure of monolayer $MoS_2$, with a direct band gap of ~1.68 eV at **K** point. Within the standard single-particle framework [9, 10], since the photon momentum is negligible on the scale of the Brillouin zone (BZ), only vertical optical transitions can occur [Fig. 2(a)], yielding the SC conductivity shown by the dashed line in Fig. 2(b). For $\hbar\omega < 2.6$ eV, two peaks of opposite sign (labeled $\delta_1$ and $\delta_2$) arise from vertical transitions between the topmost valence band ($VB_1$) and the lowest conduction band ($CB_1$) [hereafter, $CB_n$ ($VB_n$) denotes the $n$-th CB (VB) counted from the band edges]. For $\hbar\omega > 2.6$ eV, the optical transitions involving deeper bands contribute substantially, yielding a maximum $\sigma^{(2)}_{yxx}$ of ~22 nm·μA/V². Including spin-orbit coupling (SOC) splits the peaks at 1.6 eV and 3.2 eV into sub-peaks, arising from the splitting of the VB edge and of $CB_3$ at **K** point, respectively (see Secs. II and III in the SI [37]); however, the overall spectral shape remains largely unchanged. We therefore neglect SOC in the main analysis and return to its impacts in the final discussion.

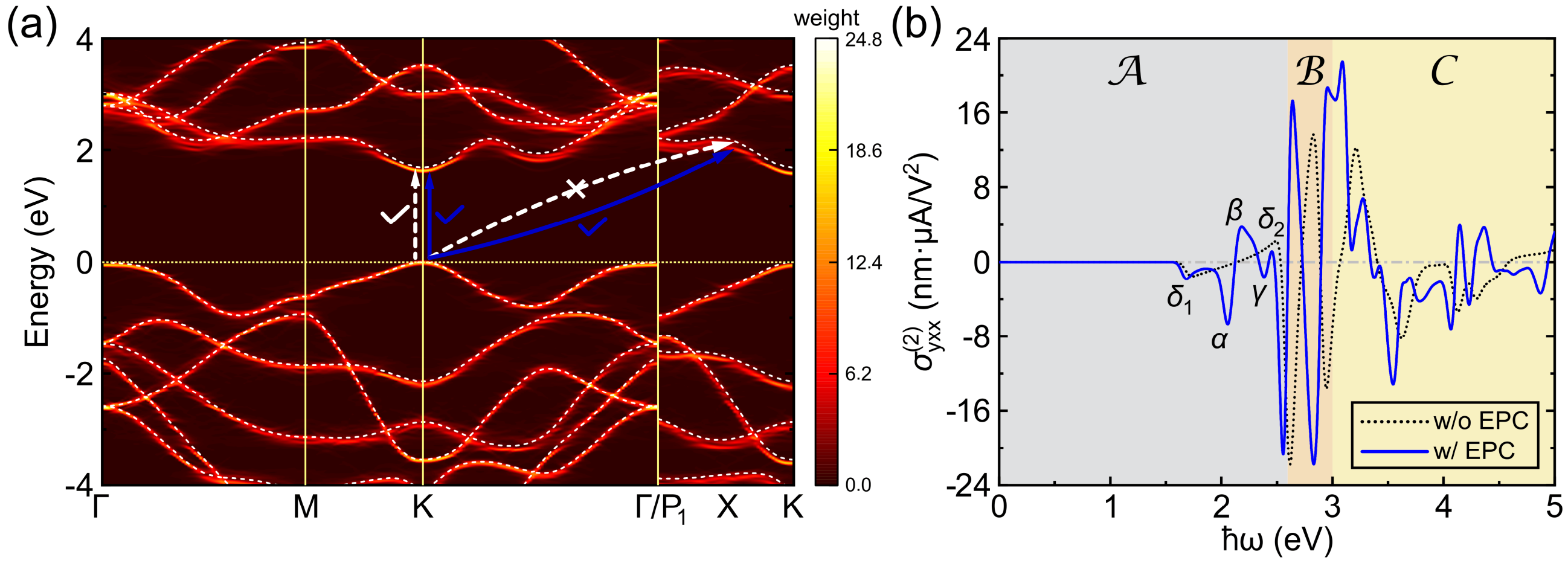

**FIG. 2.** Band renormalization and BRSC of monolayer $MoS_2$. (a) EPC-renormalized spectral function at $T = 0$ K; the bare band structure is overlaid as white dashed lines. White dashed arrows denote vertical optical transitions in the single-particle picture; blue solid arrows exemplify phonon-assisted transitions between momentum-distinct states enabled by band broadening. (b) SC conductivity $\sigma^{(2)}_{yxx}$ at $T = 0$ K with (blue solid) and without (black dashed) EPC. Three additional peaks, labeled $\alpha$, $\beta$,

and $\gamma$, appear in the low-energy region $\mathcal{A}$ alongside the single-particle peaks $\delta_1$ and $\delta_2$; regions $\mathcal{B}$ and $\mathcal{C}$ mark the redshifted and reconfigured regimes, respectively.

To assess the impacts of EPC on SC, we have calculated the EPC-renormalized spectral function of monolayer $MoS_2$ at zero temperature. As shown in Fig. 2(a), the centers of spectral weight coincide almost perfectly with the bare bands, indicating that the ZPR only slightly alters the dispersion: it reduces the band gap from 1.68 to 1.60 eV, a mere ~80 meV (~5.0% of the original gap). In contrast, the broadening of the spectral function is strongly momentum-dependent — minor at **K** and **Γ** points but pronounced for states elsewhere in the BZ. With increasing temperature, the broadening grows progressively and the band gap shrinks monotonically (see Sec. IV in the SI [37]). As we show below, it is this momentum-resolved broadening that enables otherwise forbidden optical transitions between states with different momenta [Fig. 2(a)].

Using the renormalized spectral function, we can calculate the BRSC conductivity $\sigma^{(2)}_{yxx,\mathrm{BR}}$ via the finite-difference method [40, 41]:

$$\sigma^{(2)}_{yxx,\mathrm{BR}}(T) = \sigma^{(2)}_{yxx} + \sum_{\boldsymbol{q},\nu} \frac{1}{2\omega_{\boldsymbol{q}\nu}} \frac{\partial^2 \sigma^{(2)}_{yxx}}{\partial u^2_{\boldsymbol{q}\nu}} \left[\frac{1}{2} + n_B(\omega_{\boldsymbol{q}\nu}, T)\right]. \text{(Eq. 1)}$$

Here, $T$ is temperature, $\omega_{\boldsymbol{q}\nu}$ and $u_{\boldsymbol{q}\nu}$ are the frequency and eigenvector of the phonon with momentum $\boldsymbol{q}$ and eigenmode $\nu$, respectively, and $n_B$ is the Bose–Einstein distribution function. The blue solid line in Fig. 2(b) shows $\sigma^{(2)}_{yxx,\mathrm{BR}}$ at $T = 0$ K. Strikingly, ZPR induces dramatic changes in $\sigma^{(2)}_{yxx,\mathrm{BR}}$, and the resulting spectrum falls into three distinct energy regions relative to its EPC-free counterpart. In the low-energy region (region $\mathcal{A}$, $\hbar\omega < 2.6$ eV), three entirely new peaks, labeled $\alpha$, $\beta$, and $\gamma$, appear alongside the single-particle peaks $\delta_1$ and $\delta_2$. At intermediate energies (region $\mathcal{B}$, 2.6 eV $< \hbar\omega < 3.0$ eV), the $\sigma^{(2)}_{yxx,\mathrm{BR}}$ peaks are redshifted with enhanced absolute magnitudes. In the higher-energy region (region $\mathcal{C}$, $\hbar\omega > 3.0$ eV), the spectrum deviates substantially from its bare counterpart, exhibiting a complex reconfiguration that precludes simple proportionality. These pronounced modifications originate from the EPC-induced restructuring of electronic states, a transition from infinite to finite electronic lifetimes that fundamentally alters the optical transition probabilities.

The momentum-dependent broadening identified above is quantified by the electron-phonon scattering rate, *i.e.*, the inverse quasiparticle lifetime $\tau_{n\boldsymbol{k}}$ [36]:

$$\frac{1}{\tau_{n\boldsymbol{k}}} = \frac{2\pi}{\hbar}\sum_{m\nu}\int\frac{d\boldsymbol{q}}{\Omega_{BZ}}|g_{nm\nu}(\boldsymbol{k},\boldsymbol{q})|^2$$
$$\times\left[\left(1 - f_{m\boldsymbol{k}+\boldsymbol{q}} + n_{\boldsymbol{q}\nu}\right)\delta\left(\varepsilon_{n\boldsymbol{k}} - \hbar\omega_{\boldsymbol{q}\nu} - \varepsilon_{m\boldsymbol{k}+\boldsymbol{q}}\right)\right.$$
$$\left. + \left(f_{m\boldsymbol{k}+\boldsymbol{q}} + n_{\boldsymbol{q}\nu}\right)\delta\left(\varepsilon_{n\boldsymbol{k}} + \hbar\omega_{\boldsymbol{q}\nu} - \varepsilon_{m\boldsymbol{k}+\boldsymbol{q}}\right)\right], \text{(Eq. 2)}$$

where $g_{nm\nu}(\boldsymbol{k},\boldsymbol{q})$ is the matrix element measuring EPC strength, $\varepsilon_{n\boldsymbol{k}}$ and $\hbar\omega_{\boldsymbol{q}\nu}$ are the energies of the electron and phonon, $f_{m\boldsymbol{k}+\boldsymbol{q}}$ is the Fermi–Dirac distribution function; the first and second terms in square brackets describe phonon emission and absorption, respectively. In Eq. (2), $\boldsymbol{k}$ and $\boldsymbol{q}$ label the momenta of the electron and phonon, while $n$ ($m$) and $\nu$ are the indices of energy bands and phonon eigenmodes, respectively. At $T = 0$ K, the electronic states near the band edges lack the phase space for phonon scattering, giving rise to a long lifetime, consistent with the minor broadening at **K** and **Γ** points in Fig. 2(a). To connect the scattering with optical excitations, we group the CB and VB states according to their mutual energy separation $\Delta E = \varepsilon_{\mathrm{CB}} - \varepsilon_{\mathrm{VB}}$ and plot the corresponding scattering rates in Figs. 3(a) and 3(b). Three distinct regions emerge, matching the three spectral intervals defined above: region $\mathcal{A}$ ($\Delta E < 2.6$ eV), region $\mathcal{B}$ (2.6 eV $< \Delta E <$ 3.0 eV), and region $\mathcal{C}$ ($\Delta E >$ 3.0 eV).

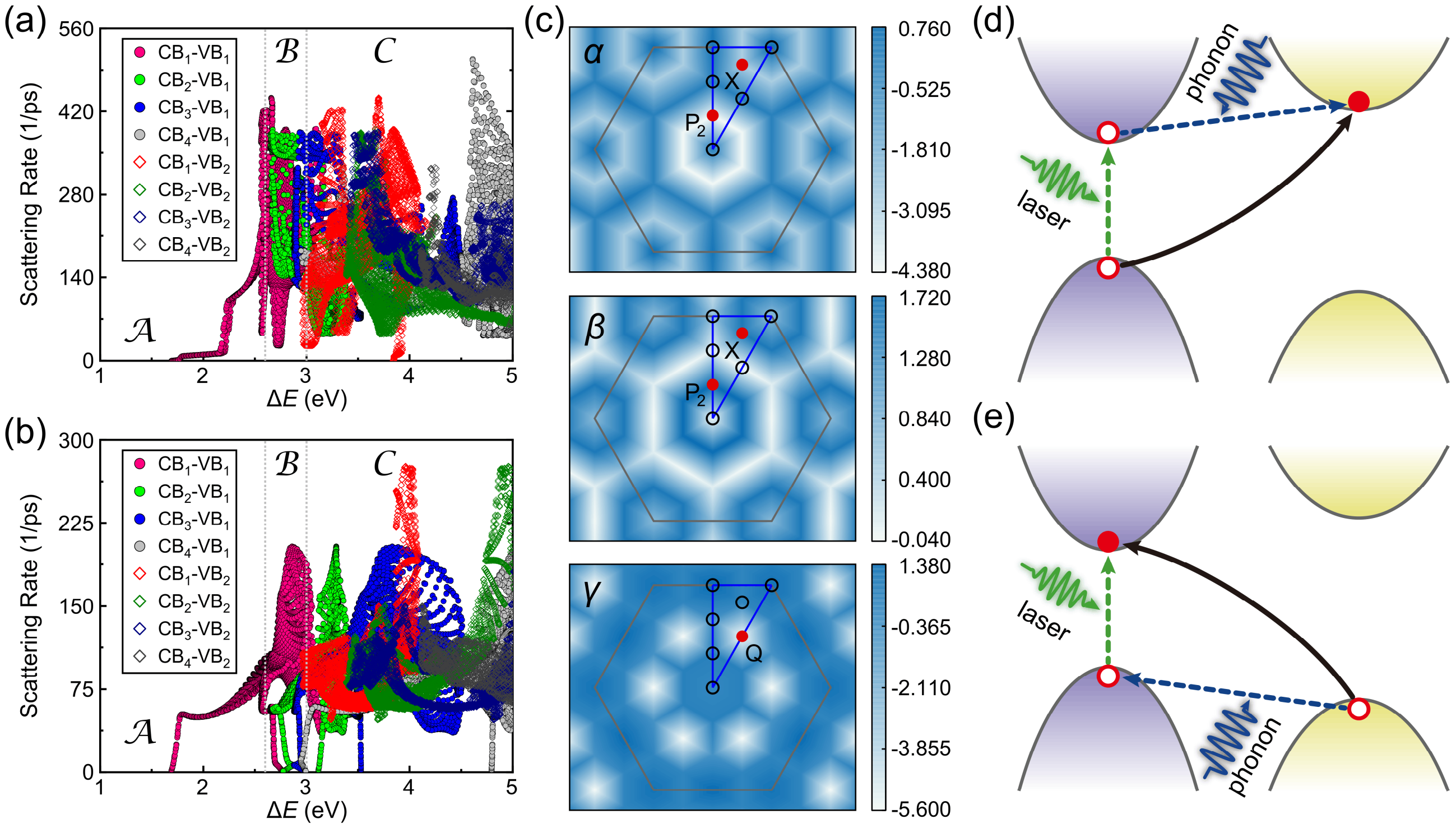

**FIG. 3.** Microscopic mechanism of BRSC. (a) The electron-phonon scattering rates for states in the conduction bands $\mathrm{CB}_n$ and (b) those in the valence bands $\mathrm{VB}_n$ at $T = 0$ K, grouped by the energy separation $\Delta E = \varepsilon_{\mathrm{CB}} - \varepsilon_{\mathrm{VB}}$ between the CB and VB states involved; $n$ labels the $n$-th band counted from the band gap. Vertical lines delimit the regions $\mathcal{A}$, $\mathcal{B}$, and $\mathcal{C}$. (c) Phonon-wavevector-resolved

contributions to $\sigma^{(2)}_{yxx,\mathrm{BR}}$ at the photon energies of peaks $\alpha$ (2.06 eV), $\beta$ (2.18 eV), and $\gamma$ (2.38 eV), mapped over the BZ; the relevant $\boldsymbol{q}$ points (**P₂**, **X**, and **Q**) are marked. (d) and (e) The two phonon-assisted pathways: (d) photon-first — vertical VB→CB excitation at **K** point followed by phonon scattering of the states in the CB; (e) phonon-first — scattering within the VB (equivalently, hole creation away from **K** point) followed by vertical excitation at **K** point.

In region $\mathcal{A}$, which by construction involves only $CB_1$ and $VB_1$ states, the scattering rates are already appreciable, so that EPC reshapes the band-edge optical response directly. In region $\mathcal{B}$, $CB_2$ states enter the scattering phase space with relatively large rates, while VB scattering remains confined to $VB_1$. In region $\mathcal{C}$, the four lowest CBs, and notably $VB_2$, all exhibit significant scattering, giving rise to highly complex multi-state processes. Since the emergent peaks $\alpha$, $\beta$, and $\gamma$ — the central finding of this work — reside in region $\mathcal{A}$ and arise from phonon-assisted $VB_1$–$CB_1$ transitions enabled solely by scattering within the $VB_1$ and $CB_1$, we focus our microscopic analysis on this region. The spectral changes in regions $\mathcal{B}$ and $\mathcal{C}$ instead reflect the intertwined competition of multiple phonon-assisted channels, which precludes a one-to-one peak assignment.

We now identify the microscopic origin of the three emergent peaks by resolving Eq. (1) with respect to the phonon wavevector $\boldsymbol{q}$. Figure 3(c) maps the resulting $\boldsymbol{q}$-resolved contribution to $\sigma^{(2)}_{yxx,\mathrm{BR}}$ across the BZ at the peak photon energies, *i.e.*, 2.06, 2.18, and 2.38 eV for peaks $\alpha$, $\beta$, and $\gamma$, respectively. Peaks $\alpha$ and $\beta$ share the same set of mediating phonons, with wavevectors $\boldsymbol{q}$ = $\mathbf{P}_2$ [(0, 1/6, 0) and symmetry equivalents, along **Γ**–**M**] and $\boldsymbol{q}$ = **X** [(1/6, 1/3, 0) and symmetry equivalents, low-symmetry $\boldsymbol{q}$-points]; their opposite signs instead originate from the different shift vectors of the underlying transition paths. Peak $\gamma$, by contrast, is dominated by phonons at $\boldsymbol{q}$ = **Q** [(1/6, 1/6, 0) and symmetry equivalents, along **Γ**–**K**], with negligible contributions from other $\boldsymbol{q}$ points.

The formation of each emergent peak can be traced to two distinct phonon-assisted transition pathways, as illustrated in Figs. 3(d) and 3(e). In the first pathway [Fig. 3(d)], a photon first drives a vertical $VB_1$→$CB_1$ transition at **K** point, and the excited electron is then scattered within $CB_1$ from **K** to **X** by phonons with $\boldsymbol{q}$ = $\mathbf{P}_2$/**X** or equivalents (see Fig. 3(c) and Sec. V in the SI [37]). The net process is a phonon-assisted transition between $VB_1$ at **K** and $CB_1$ at **X**, with $\Delta E \approx$ 2.20 eV, contributing to peak $\beta$. In the second pathway [Fig. 3(e)], the phonon acts first: an electron is scattered within $VB_1$ from **X** to **K** (equivalently, a hole from **K** to **X**) by phonons with $\boldsymbol{q}$ = $\mathbf{P}_2$/**X** or equivalents (see Fig. 3(c) and Sec. V in the SI [37]), and the vertical optical excitation follows at **K** point. This amounts to a phonon-assisted transition between $VB_1$ at **X** and $CB_1$ at **K**, with $\Delta E \approx 2.15$ eV, likewise feeding peak $\beta$. We emphasize that neither channel exists without EPC [see the vertical-only arrows in Fig. 2(a)]. Finally, we note that peak $\beta$ acquires its finite width from the large broadening of the states near **X** involved in both pathways, while the **K**-side states remain sharp — consistent with the momentum-resolved broadening in Fig. 2(a).

The same analysis applies to peaks $\alpha$ and $\gamma$. For peak $\alpha$, the electrons initially near the $\mathbf{P_2}$ points in $VB_1$ are scattered to **K** by phonons with $\boldsymbol{q}$ = **X** or equivalents (see Fig. 3(c) and Sec. V in the SI [37]) and then excited vertically to $CB_1$ at **K** point; the net $\mathbf{P_2}\rightarrow\mathbf{K}$ transition has $\Delta E \approx 1.97$ eV. For peak $\gamma$, the electrons near the **M** (**Q**) points in $VB_1$ undergo **M**→**K** (**Q**→**K**) scattering assisted by phonons with $\boldsymbol{q}$ = **Q** or equivalents (see Fig. 3(c) and Sec. V in the SI [37]), followed by vertical excitation to $CB_1$ at **K** point, with $\Delta E \approx 2.30$ (2.45) eV. Different from peak $\beta$, peaks $\alpha$ and $\gamma$ are dominated by the phonon-first pathway alone. It worth noting that the observed peak maxima at 2.06 and 2.38 eV lie within the spread of the contributing transition energies, reflecting the combined spectral weight of these channels.

In regions $\mathcal{B}$ and $\mathcal{C}$, the number of participating bands, and hence of phonon-assisted channels, grows rapidly with the energy separation $\Delta E$ [Figs. 3(a) and 3(b)], so that a one-to-one assignment of peaks to individual processes is no longer meaningful. The observed systematics nevertheless follow directly from the channel statistics: in region $\mathcal{B}$, the proliferation of scattering channels strengthens the phonon-assisted response while the channels remain few enough to retain a well-defined peak structure, accounting for the redshifted and enhanced peaks in Fig. 2(b). In region $\mathcal{C}$, the channels involve states far from the band edges whose individual optical matrix elements are suppressed, and their mutual competition flattens the spectrum; the $\sigma^{(2)}_{yxx,\mathrm{BR}}$ accordingly remains modest and loses any simple correspondence with the bare SC.

Because the emergent peaks originate from transitions involving states away from **K** — whose orbital characters, and hence deformation potentials, differ markedly from those of the band-edge states [42] — their spectral positions respond sensitively and non-uniformly to lattice strain. As shown in Fig. 4(a), varying the biaxial strain $\varepsilon$ from +1.0% (tensile) to −1.0% (compressive) shifts the $\sigma^{(2)}_{yxx,\mathrm{BR}}$ spectrum by ~0.2 eV; a mere 1.0% compressive strain raises $\Delta E$ enough to merge peak $\gamma$ into the adjacent higher-energy features. Such strain levels are experimentally realistic, as the atomically thin films on substrates routinely experience lattice-mismatch strain of this magnitude. Since practical photodetection operates at fixed laser frequencies, a strain-induced shift of this size can drastically alter the $\sigma^{(2)}_{yxx,\mathrm{BR}}$ intensity at a given photon energy [see the inset of Fig. 4(a)]. The bare SC, by contrast, varies smoothly between two $\delta$ peaks, so the same shift leaves it nearly unchanged (see the inset of Fig. 4(a) and Sec. VI in the SI [37]). These strain-sensitive intensity variations may help reconcile persistent discrepancies between theoretical SC predictions and experimental measurements [43, 44].

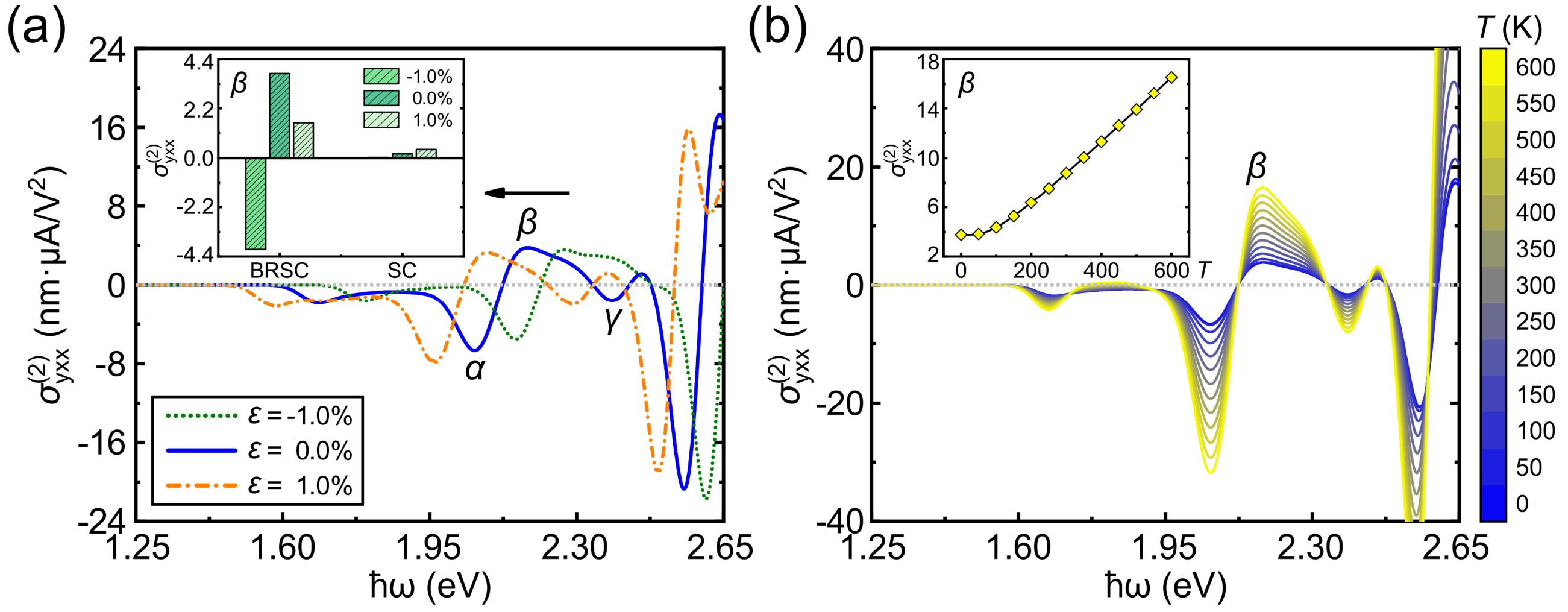

**FIG. 4.** Tunability of the BRSC under external stimuli. (a) $\sigma^{(2)}_{yxx,\mathrm{BR}}$ conductivity under biaxial strain $\varepsilon$; the spectrum shifts by ~0.2 eV over this range. Inset: comparison between $\sigma^{(2)}_{yxx,\mathrm{BR}}$ and bare $\sigma^{(2)}_{yxx}$ at a fixed photon energy of 2.18 eV for different values of $\varepsilon$. (b) $\sigma^{(2)}_{yxx,\mathrm{BR}}$ conductivity at temperatures from 0 to 600 K. Inset: the $\sigma^{(2)}_{yxx,\mathrm{BR}}$ for peak $\beta$ versus temperature, showing an enhancement by a factor of ~4.4.

Different from strain, which tunes the transition energy $\Delta E$, temperature acts directly on the phonon population: increasing $T$ enhances the electron-phonon scattering and activates otherwise forbidden transition channels, leading to an overall enhancement of the BRSC. Figure 4(b) presents $\sigma^{(2)}_{yxx,\mathrm{BR}}$ from 0 to 600 K; peak $\beta$, for instance, grows by a factor of ~4.4. Since all temperature dependence enters solely via phonon occupation $n_B$, the enhancement follows an approximately parabolic temperature dependence at low temperature [inset of Fig. 4(b)], which arises from the combined contributions of phonons with nonvanishing coupling (long-wavelength-limit acoustic modes do not contribute). At high temperature, the response gradually crosses over to linear temperature behavior as $n_B$ scales linearly with temperature [40, 41]. We note that at elevated temperatures phonon-phonon interactions become important and push the system beyond the regime of validity of Eq. (1); the impact of anharmonicity on the BRSC thus deserves further investigation.

*Discussion.* We now return to SOC neglected so far. In monolayer group-VI TMDs, SOC splits the SC peaks into sub-peaks — for monolayer $MoS_2$, at $\hbar\omega = 1.6$ eV ($VB_1$-derived splitting, ~150 meV at **K** point) and 3.2 eV ($CB_3$-derived splitting, ~140 meV at **K** point) (see Secs. II and III in the SI [37]). At zero temperature, where the EPC broadening at **K** remains minor, these sub-peaks in principle persist in the BRSC. At finite temperature, however, the broadening grows steadily (see Sec. IV in the SI [37]) and becomes comparable to or larger than these splittings — most readily for the $CB_3$-derived splitting — so that the sub-structures are smeared out and become inconspicuous in the BRSC spectrum (see Secs. VII and VIII in the SI [37]).

Finally, we discuss the relation to excitonic effects, the other prominent many-body correction in monolayer TMDs [45–50]. Since the exciton binding energy in monolayer $MoS_2$ exceeds the ZPR of ~80 meV, excitonic enhancement of the SC is undeniably important [26–30], and we set it aside here only to isolate the EPC contribution. The two corrections are, however, experimentally distinguishable: excitons introduce discrete resonances near the absorption edge and enhance existing SC peaks, whereas the BRSC produces new peaks from momentum-forbidden transitions, a redshift of higher-energy features, and — most distinctively — a monotonic nonlinear temperature enhancement rooted in the phonon occupation. Recent experiments on twisted bilayer $MoS_2$ reporting phonon-assisted signatures in the BPVE [51] suggest that such an EPC contribution is already within experimental reach. A unified treatment of both many-body effects remains an open challenge and a natural next step. In bulk polar materials with strong dielectric screening, where excitonic effects are weak, the EPC-driven band renormalization identified here should constitute the dominant many-body correction to the SC.

*Conclusion.* We have established the band-renormalized shift current (BRSC) — the contribution of electron-phonon coupling (EPC) to the shift current (SC) via virtual-phonon band renormalization — in monolayer $MoS_2$ from first principles. While a zero-point renormalization barely alters the band dispersion, it completely restructures the SC spectrum even at zero temperature. Moreover, the BRSC is highly tunable via strain because of the deformation-potential contrast of the states involved, and raising the temperature enhances the BRSC. These fingerprints — momentum-forbidden peaks and monotonic thermal growth — offer experimentally testable signatures that distinguish the EPC channel from excitonic corrections and may help resolve long-standing discrepancies between calculated and measured nonlinear photocurrents. Because the underlying mechanism requires only a non-centrosymmetric structure and EPC — both of which are ubiquitous, we expect band renormalization to be a general ingredient of nonlinear photocurrents across polar materials.

*Acknowledgement.* We acknowledge Drs. Shengnan Xu, and Xiao Jiang for helpful discussions. This work is supported by National Natural Science Foundation of China (NSFC) (Grants Nos. 12347128, W2511008, 12404260) and Science Challenge Project (Grant No. TZ2025013).